\documentclass{article}
\usepackage{spconf,amsmath,graphicx}
\usepackage{soul}
\usepackage{threeparttable}
\usepackage{multirow}
\usepackage{booktabs}
\usepackage{cite}
\usepackage{url}

\title{Self-Transriber: Few-shot Lyrics Transcription with Self-training}
\name{Xiaoxue Gao$^1$, Xianghu Yue$^1$, Haizhou Li$^{1,2,3}$}
\address{
  $^1$Department of Electrical and Computer Engineering, National University of Singapore, Singapore \\
  $^2$Shenzhen Research Institute of Big Data, Shenzhen, China \\
  $^3$School of Data Science, The Chinese University of Hong Kong, Shenzhen, China\\
\thanks{
\textit{Corresponding author: Xianghu Yue}.
}
}

\begin{document}
%
\maketitle

\begin{abstract}
\vspace{-0.2cm}
The current lyrics transcription approaches heavily rely on supervised learning with labeled data, but such data are scarce and manual labeling of singing is expensive. How to benefit from unlabeled data and alleviate limited data problem have not been explored for lyrics transcription. We propose the first semi-supervised lyrics transcription paradigm, Self-Transcriber, by leveraging on unlabeled data using self-training with noisy student augmentation. We attempt to demonstrate the possibility of lyrics transcription with a few amount of labeled data. 
Self-Transcriber generates pseudo labels of the unlabeled singing using teacher model, and augments pseudo-labels to the labeled data for student model update with both self-training and supervised training losses.
This work closes the gap between supervised and semi-supervised learning as well as opens doors for few-shot learning of lyrics transcription. 
Our experiments show that our approach using only 12.7 hours of labeled data achieves competitive performance compared with the supervised approaches trained on 149.1 hours of labeled data for lyrics transcription.
\end{abstract}
\vspace{-0.2cm}
\begin{keywords}
Lyrics transcription, music information retrieval, singing processing and self-training.
\end{keywords}
\vspace{-0.4cm}
\section{Introduction}
\label{sec:intro}
\vspace{-0.4cm}
Singing helps evoke and convey human emotions, and understanding sung text contributes to the perception of songs and helps listeners’
enjoyment of singing~\cite{fine2014making,casey2008content,gao2019speaker,gao2020personalized}.
Automatic lyrics transcription (ALT) aims to understand their relationship by transcribing lyrical text from singing vocals. ALT has facilitated many applications in music information retrieval such as automatic generation of karaoke lyrical content, music video subtitling, and query-by-singing \cite{mesaros2013singing,gao2022music,mesaros2010automatic,dzhambazov2015modeling,gao2022genre}. 

Despite the rapid development of automatic speech recognition (ASR)~\cite{sun2019adversarial,liu2021tera,baevski2021unsupervised,yue2022}, ALT has not been well studied due to the lack of publicly available data and its inherent peculiarities originally existing in
singing. Past studies have focused on two aspects for ALT: 1) the development of databases with paired singing vocal and lyrics labels and 2) the exploration of supervised learning techniques including Kaldi-hybrid models \cite{gupta2019,mesaros2010automatic,dzhambazov2015modeling} and end-to-end models~\cite{gao2022automatic,ou2022towards}. 

Data sparsity issue limits the advance of lyrics transcription. Thanks to the increasing attention on music information retrieval (MIR)~\cite{kim2020semantic}, the solo-singing karaoke dataset DAMP~\cite{damp_early} was released by Smule in 2017, followed by an updated version \cite{DAMPDataLatest} with time-aligned lyrics information and more songs from 30 countries. Dabike et al.~\cite{dabike2019automatic} further processed the data and created a cleaner version of DAMP Sing! 300x30x2 (Sing!)~\cite{DAMPDataLatest}, thereby offering more possibilities for lyrics transcription.

Thanks to the increase of the publicly available datasets~\cite{damp_early,DAMPDataLatest,dabike2019automatic}, it is feasible to incorporate data-driven supervised learning approaches for ALT.
For example, supervised approaches using Kaldi-hybrid architecture~\cite{povey2011kaldi} are introduced including DNN-HMM~\cite{gupta2018semi, gupta2018automatic}, factorized Time-Delay Neural Network (TDNN-F)~\cite{dabike2019automatic}, and self-attention based TDNN-F acoustic models~\cite{demirel2020automatic}.
Moreover, E2E supervised models using transformer~\cite{gao2022automatic} and transfer learning \cite{ou2022towards} are also explored for ALT.
Though these supervised learning approaches successfully achieve automatic lyrics transcription of solo-singing, they are trained in a supervised fashion, which heavily relies on having a dataset that provides both the singing audio and its corresponding lyrical labels to succeed.  

\begin{figure*}[t]
\centering
\vspace{-1cm}
\includegraphics[width=161mm]{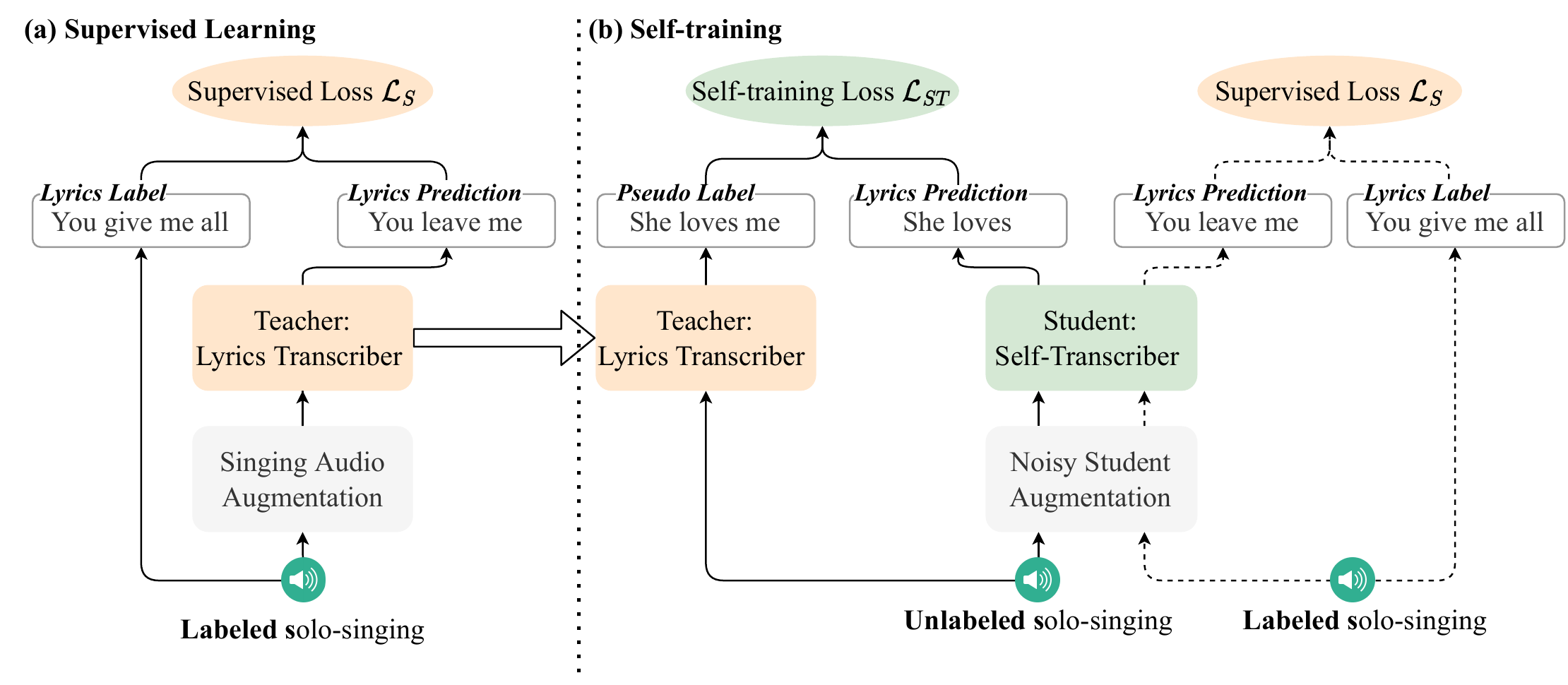}
\vspace{-0.7cm}
\caption{The network architecture and training flow of the proposed Self-Transcriber with (a) supervised learning and (b) self-training techniques.}
\label{main}
\vspace{-0.6cm}
\end{figure*}

Even if some databases~\cite{damp_early,DAMPDataLatest,dabike2019automatic} have been made publicly available, large-scale data labeling of singing is still a very laborious and challenging task, making it not a sustainable method to use in the future.
Specifically, humans with normal hearing ability and language understanding can recognize words while listening to speech, but they often have difficulty in recognizing sung lyrics from solo-singing~\cite{fine2014making}. Lyrics transcription of solo-singing is challenging even for professional musicians~\cite{fine2014making}.
Therefore, how to get rid of manual labeling and make use of unlabeled data becomes essential for ALT.

Knowing that there are a large amount of languages and dialects existing in the world, it's difficult and expensive to collect hundreds of hours of labeled singing data for most of them for the development of supervised lyrics transcription. This creates the expectation that the current ALT techniques can be further improved without obtaining a large amount of labeled data to perform well, which is the focus of this paper.

Inspired by the success of self-training models in natural language processing\cite{du2021self,chen2021revisiting}, ASR~\cite{chen2020semi,kahn2020self,xu2021self} and computer vision~\cite{ghiasi2021multi}, we propose a Self-Transcriber framework by benefiting from unlabeled data using self-training technique to resolve the data sparsity problem in lyrics transcription. Self-training is one important training approach of semi-supervised learning techniques~\cite{chen2020semi,kahn2020self,xu2021self}.
Our contributions include the following: a) we propose a novel semi-supervised ALT framework by benefiting from unlabeled data through self-training technique for the first time; b) we investigate noisy student augmentation technique for self-training of ALT; c) our solution achieves competitive performance on few-shot learning of lyrics transcription using only 12.7h labeled data compared with current SOTA supervised approaches using  149.1 labeled data, which provides flexibility in the training requirements by removing the need for a large amount of labeled data to perform well.

\vspace{-0.3cm}
\section{Self-transcriber}
\vspace{-0.3cm}
Though supervised training enables the recognition of lyrics from solo-singing, it is impractical to collect large-scale labeled singing data from a large majority of languages for purely supervised learning. Therefore, we study self-training by leveraging unlabeled data and facilitate few-shot lyrics transcription. 

We propose a Self-Transcriber model that includes lyrics transcriber teacher modeling in a supervised manner as detailed in Fig.~\ref{main} (a), followed by the student lyrics transcriber modeling using self-training in Fig.~\ref{main} (b). We employ self-training by pseudo-labeling the unannotated data using a trained teacher model and then retrain the lyrics transcriber with the additional labeled data. 
\vspace{-0.2cm}
\subsection{Supervised Training}
\label{Supervised Training}

We begin by training a supervised model for ALT, that we call as teacher lyrics transcriber in Fig.~\ref{main} (a).
Moreover, E2E supervised models using transformer~\cite{gao2022automatic} and transfer learning \cite{ou2022towards} are also explored for ALT.
Lyrics transcriber takes the solo-singing signal as input and generates lyrics as output, and it is based on a joint encoder-decoder architecture including wav2vec2.0 feature extractor and encoder together with an attention-based GRU decoder~\cite{baevski2020wav2vec,ou2022towards,gu2022mm}. Lyrics transcriber first converts augmented solo-singing audio~\cite{park2019specaugment} into solo-singing representation via wav2vec2.0, and then GRU decoder predicts lyric characters one at a time given the solo-singing representations and the previously predicted lyrical tokens in an autoregressive manner. A multi-task supervised objective function $\mathcal{L}_{\text{S}}$ is employed including connectionist temporal classification (CTC) loss $\mathcal{L}_{\text{CTC}}$ and sequence-to-sequence (S2S) loss $\mathcal{L}_{\text{S2S}}$ between lyrics prediction and true labels for supervised model training following~\cite{ou2022towards,gao2022automatic}.
\vspace{-0.15cm}
\begin{equation}
\vspace{-0.1cm}
     \mathcal{L}_{\text{S}} = \beta \mathcal{L}^{\text{CTC}}_{\text{S}} + (1-\beta) \mathcal{L}^{\text{S2S}}_{\text{S}}
\label{Sloss}
\end{equation}
where $\beta \in [0,1]$ serves as a multi-task interpolation factor. We take the trained supervised model as the first teacher model for self-training. 
\vspace{-0.4cm}
\subsection{Semi-supervised Lyrics Transcription}
\vspace{-0.2cm}
Inspired by the fact that semi-supervised models can benefit from partially labeled data \cite{chen2021revisiting,chen2022generate}, we propose the first semi-supervised lyrics transcription model, Self-Transcriber, that is trained on both labeled and unlabeled data using self-training and data augmentation paradigms, as illustrated in Fig.~\ref{main} (b).

\subsubsection{Noisy Student Augmentation on Unsupervised Data}

Motivated by the succeed of data augmentation \cite{chen2020semi,chen2021revisiting} and noisy student \cite{xie2020self} for self-training, we propose to perform noisy student augmentation on the unsupervised data by using pseudo lyrical labels decoded from the original solo-singing data on its distorted audio versions. During pseudo labeling, the teacher is not noised to retain good pseudo labels, but the student is noisy through performing the singing audio augmentation on solo-singing audio by speed perturbation, dropping frequency bands and masking blocks of time steps~\cite{park2019specaugment}. To this regard, the student is generalized better than the teacher.
\vspace{-0.4cm}
\subsubsection{Self-training for Lyrics Transcription}
\vspace{-0.2cm}
We propose to adopt self-training on the unsupervised data for lyrics transcription of solo-singing.
True lyrics label is used for supervised data and self-training is employed by pseudo labeling unsupervised solo-singing data through teacher model together with the pretrained language model (LM). Self-Transcriber first performs student noisy augmentation on both labeled and unlabeled data, and then augmented data goes through wav2vec2.0 component and decoder following Section~\ref{Supervised Training} to predict lyrics. Self-training loss $\mathcal{L}_{\text{ST}}$ is calculated between predicted lyrical tokens and pseudo labels as below.
\begin{equation}
\vspace{-0.1cm}
     \mathcal{L}_{\text{ST}} = \beta \mathcal{L}^{\text{CTC}}_{\text{ST}} + (1-\beta) \mathcal{L}^{\text{S2S}}_{\text{ST}}
\vspace{-0.1cm}
\label{ST_loss}
\end{equation}
where $\mathcal{L}^{\text{CTC}}_{\text{ST}}$ and $\mathcal{L}^{\text{S2S}}_{\text{ST}}$ are self-training based CTC and S2S losses, respectively. To explore the proportional contributions between the two objectives, the unified loss of Self-Transcriber $\mathcal{L}$ is formulated to consider both supervised loss $\mathcal{L}_{\text{S}}$ and self-training loss $\mathcal{L}_{\text{ST}}$ with semi-supervised factor $\alpha$ included.
\begin{equation}
\vspace{-0.2cm}
    \mathcal{L} =\mathcal{L}_{\text{S}}+ \alpha\mathcal{L}_{\text{ST}}
\label{alpha}
\vspace{-0.1cm}
\end{equation}
where supervised-training loss $\mathcal{L}_{\text{S}}$ is applied for labeled data as in Eq.~\ref{Sloss}, and semi-supervised factor $\alpha \in [0,1]$ is introduced to accommodate label noise introduced by self-training technique.

After the first iteration of self-training as in Fig.~\ref{main} (b), we propose to re-generate lyrical pseudo label using trained Self-Transcriber and then re-train a new Self-Transcriber model with labeled data and updated pseudo labels, so as to have the two reinforce each other, and to continuously improve both. Note that the weights of wav2vec2.0 feature extractor and transformer are initialized by the model pretrained and finetuned on speech data for all supervised training and self-training models.

\vspace{-0.5cm}
\section{Experiments}
\vspace{-0.5cm}
\subsection{Dataset}
\vspace{-0.2cm}
A English solo-singing dataset~\cite{dabike2019automatic} $\textit{Sing! 300} \times \textit{30} \times \textit{2}$~\footnote{The audio files can be accessed from https://ccrma.stanford.edu/damp/} is adopted, and the details of labeled and unlabeled data are listed in Table~\ref{tab:datasets-mono}. The training set DSing1 (DS1) is adopted as labeled data and it contains 434 songs with 8,550 utterances, whose duration with more than 28s are filtered out to avoid out-of-memory issue. The development set and the test set contain 66 songs and 70 songs with 482 and 480 audio lines, respectively. The unlabeled data DS31 and DS301 are created from DSing3 (DS3) and DSing30 (DS30), respectively~\cite{dabike2019automatic}. The DS30 and DS3 sets contain 4,324 songs and 1,343 songs with a total duration of 149.1 and 44.7 hours, respectively. To keep unlabeled data and labeled data non-overlapped, utterances that are overlapped with DS1 are removed from DS3 to construct DS31 (23.3 hours), and utterances that are overlapped with DS1 or DS3 are filtered from DS30 for creating DS301 (109.3 hours). We also filter out the utterances whose duration are more than 28s for DS31 and DS301.

\begin{table}[t]
\vspace{-0.4cm}
\footnotesize
\centering
\caption{A description of the solo-singing dataset.}
\label{tab:datasets-mono}
\begin{tabular}{l|rrr}
\toprule
\textbf{Labeled Data} & \textbf{\# songs} & \textbf{\# utterances}                                                   & \textbf{duration} \\ \midrule
DS1         &    434   & 8,550 & 12.7 hours             \\ 
dev           & 66         &  482   & 0.7 hours               \\ 
test          & 70          &  480     & 0.8 hours               \\\midrule
\textbf{Unlabeled Data} & \textbf{\# songs} & \textbf{\# utt}& \textbf{duration}  \\\midrule
DS31        &     911    &   16,279 & 23.3 hours               \\ 
DS301     &   3,890    &  70,394   &  109.3 hours               \\
\bottomrule
\end{tabular}
\vspace{-0.5cm}
\end{table}

\vspace{-0.2cm}
\subsection{Experimental Setup}
\vspace{-0.2cm}
SpeechBrain toolkit~\cite{ravanelli2021speechbrain} and wav2vec2.0~\cite{baevski2020wav2vec} are used to build lyrics transcriber, which contains wav2vec2.0 with a 1024-dim linear layer followed by a leaky ReLU and a GRU deocoder~\cite{chorowski2015attention,ou2022towards}. Wav2vec2.0 mainly consists of seven CNN blocks for feature extractor and twelve transformer blocks as feature encoder~\cite{baevski2020wav2vec}. The GRU decoder is attention-based decoder with a single layer of 1,024 dimension and location-aware attention~\cite{bahdanau2016end}. All audio samples are downsampled to 16k Hz with mono-channel. All models are trained using adam optimizer~\cite{kingma2014adam} with Newbob
technique for learning rate adjusting, batch size of 2, and 10 epochs as in~\cite{ou2022towards}. We use the pre-trained Wav2vec2.0 model~\footnote{\url{https://huggingface.co/facebook/wav2vec2-large-960h-lv60-self}} for initialization and pre-trained LM~\footnote{\url{https://github.com/guxm2021/ALT_SpeechBrain/tree/main/DSing/LM}} for pseudo labeling and decoding. Hyper-parameter $\beta$ is set to 0.2 during training and beam size is set to 512 during decoding. Supervised model is trained on labeled data DS1, and four Self-Transcriber models are introduced including Self-Transcriber-S, Self-Transcriber-S2, Self-Transcriber, and Self-Transcriber2. Self-Transcriber-S and Self-Transcriber-S2 are trained on less amount of unlabeled data DS31 and labeled data DS1 with one iteration and two iterations, respectively. Self-Transcriber and Self-Transcriber2 are trained on DS1 and unlabeled data DS301 with one iteration and two iterations, respectively.

\vspace{-0.4cm}
\section{Results and Discussion}
\vspace{-0.3cm}
We study the effect of different semi-supervised factor, different amount of unlabeled data and the utilization of data augmentation to better understand self-training for lyrics transcription of solo-singing. We also conduct ablation study and compare the proposed models with state-of-the-art systems on ALT. We report the lyrics transcription performance in terms of the word error rate (WER), which is the ratio of the total number of insertions, substitutions, and deletions with respect to the total number of words. 
\vspace{-0.4cm}
\subsection{Selection of Semi-supervised Factor}
\vspace{-0.2cm}
To reflect the proportional contributions between supervised learning and self-training, we evaluate different $\alpha$ (0, 0.1, 0.9 and 1) in Eq.~\ref{alpha} on Self-Transcriber-S2 trained on labeled data DS1 and unlabeled DS31, and report their lyrics transcription in Fig.~\ref{alpha}. We observe that model without self-training performs the worst which indicates the advantage of utilizing self-training, and $\alpha$ is empirically set to 1 based on best performance of the development set, which is used for all other experiments hereafter. 
\begin{figure}[t]
\vspace{-0.7cm}
\footnotesize
\centering
\includegraphics[width=76mm]{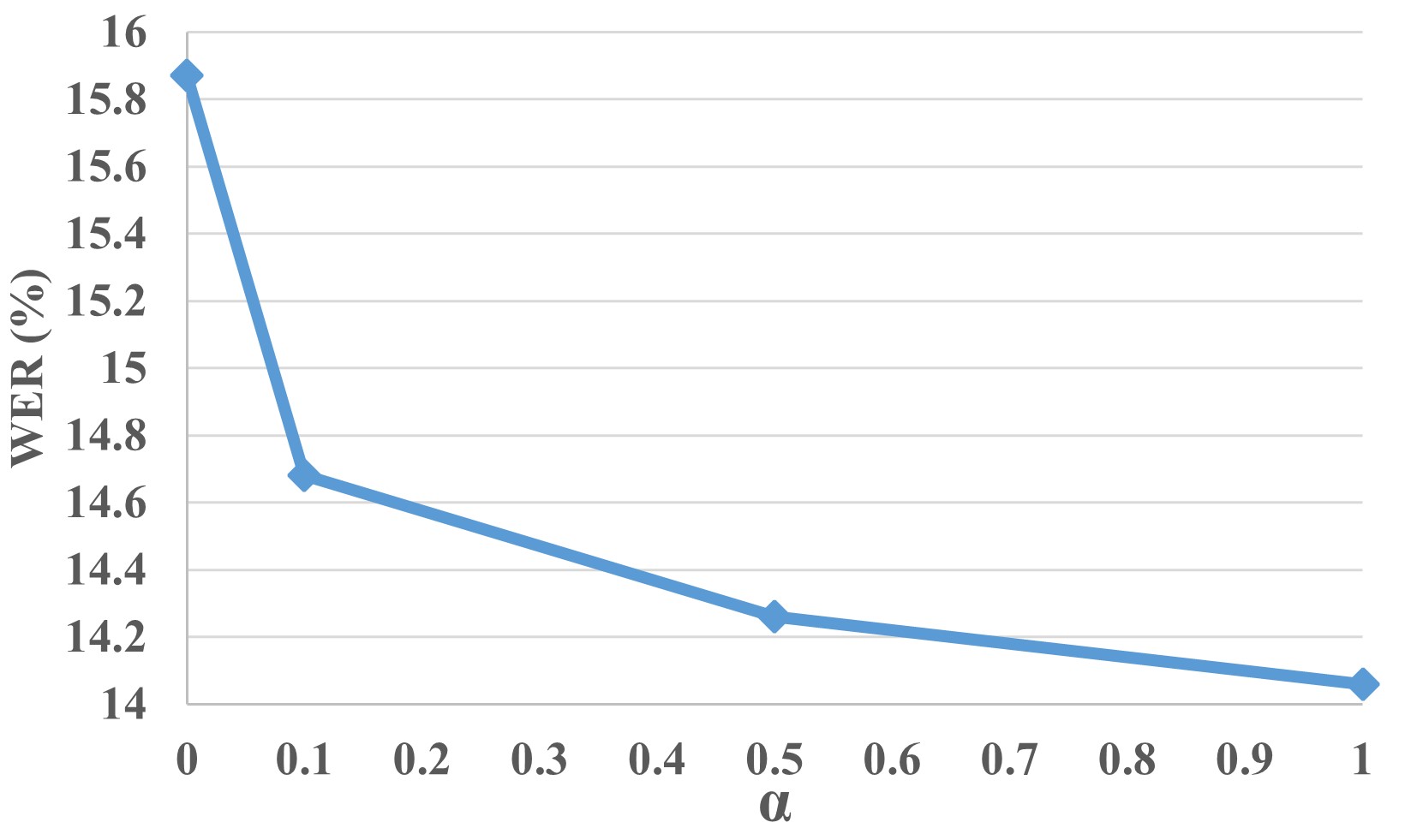}
\vspace{-0.4cm}
\caption{Comparison of lyrics recognition (WER\%) performance of Self-Transcriber-S2 with different semi-supervised factor $\alpha$ for training. Development set is used for evaluation.}
\label{alpha}
\vspace{-0.4cm}
\end{figure}

\vspace{-0.4cm}
\subsection{Effectiveness of Self-Training}
\vspace{-0.2cm}
To explore the effectiveness of self-training, we present the comparison of lyrics transcription performances on the proposed Self-Transcriber models and the supervised model baseline in Table.~\ref{mainre}. We can observe that Self-Transcriber models outperform supervised model with different amount of unsupervised data scenarios and Self-Transcriber2 model achieves 12\% relative improvement over baseline, which confirms the effectiveness of self-training on lyrics transcription. We can see that Self-Transcriber-S2 performs better than Self-Transcriber-S, which may suggest that more iterations of pseudo labeling is beneficial to self-training for ALT.
\begin{table}[t]
\footnotesize
\caption{Comparison of lyrics recognition (WER\%) performances on Self-Transcriber models with supervised model.}
\centering
\begin{tabular}{lcccc}
\toprule
\textbf{Models}   & \textbf{unlabel}   &\textbf{label}   & \textbf{dev}   & \textbf{test}  \\ \toprule
\textit{Supervised}    & 0 &DS1    &15.87&15.21  \\  \midrule
\textbf{23.3h unlabeled}   \\  
Self-Transcriber-S    & DS31 &DS1    &14.51&15.19  \\
Self-Transcriber-S2   & DS31 &DS1    &14.06&14.47  \\ \midrule
\textbf{109.3h unlabeled}   \\  
Self-Transcriber     & DS301 &DS1    &\textbf{12.39}& 13.55   \\ 
Self-Transcriber2    & DS301 &DS1    &12.86& \textbf{13.37}  \\ 
\bottomrule
\end{tabular}
\label{mainre}
\vspace{-0.4cm}
\end{table}

\vspace{-0.4cm}
\subsection{Ablation Study}
\vspace{-0.2cm}
To study where the contributions come from and verify the effectiveness of self-training and singing audio augmentation, we conduct an ablation study on the proposed Self-Transcriber2 model in Table~\ref{abl}. We observe that the removal of second iteration of self-training results in slight performance degradation for lyrics transcription.
A clear benefit of noisy student augmentation on unsupervised data can be noticed while removing second iteration and augmentation (NoisyAug). We can further observe that the removal of self-training causes a big performance drop, which verifies the effectiveness of self-training. While further removing singing audio augmentation on supervised data (SupAug), the lyrics transcription performance degrades, which highlights the beneficial of data augmentation.

\begin{table}[t]
\vspace{-0.9cm}
\footnotesize
\caption{Ablation study on the proposed Self-Transcriber with different components removed w.r.t lyrics recognition (WER\%) performances. Symbol "$-$" is removal operation.}
\centering
\begin{tabular}{l|cc|cc}
\toprule
\textbf{Models}   & \textbf{unlabel}   &\textbf{label}   & \textbf{dev}   & \textbf{test}  \\ \toprule
Self-Transcriber2     & DS301  &DS1    &12.86& \textbf{13.37}   \\ 
 \midrule 
 $-$ iteration & DS301  &DS1   &12.39& 13.55  \\  
$-$ iteration $-$ NoisyAug    & DS301  &DS1    &13.83 &14.58  \\  
$-$ self-training  & 0    &    DS1  & 15.87  &   15.21\\ 
$-$ self-training $-$ SupAug    & 0  &DS1    &16.42 &16.57  \\ 
\bottomrule
\end{tabular}
\label{abl}
\end{table}

\begin{table}[t]
\vspace{-0.6cm}
\footnotesize
\caption{Comparison between the proposed self-training solution and other existing supervised solutions on lyrics transcription performance (WER\%) of solo-singing.}
\centering
\begin{tabular}{l|c|cc}
\toprule
\textbf{Supervised models}    &\textbf{labeled data}   & \textbf{dev}   & \textbf{test}  \\ \toprule
TDNN~\cite{dabike2019automatic}  & DS30 (149.1h)    & 23.33& 19.60 \\
CTDNN-SA~\cite{demirel2020automatic} & DS30 (149.1h)    & 18.74&  14.96\\
MSTRE-Net~\cite{ahlback2021mstre}  &DS30 (149.1h)   &- & 15.38 \\ 
E2E-trans~\cite{gao2022automatic} &DS30 (149.1h)   & 13.96 &13.46  \\ 
TransferLearn~\cite{ou2022towards} &DS30 (149.1h)   &12.34 &12.99 \\
TransferLearn~\cite{ou2022towards} &Part DS30 (10h)   &- &14.84 \\\midrule
\textbf{Self-training model}    &\textbf{labeled data}   & \textbf{dev}   & \textbf{test}  \\ \midrule
Self-Transcriber2      &DS1 (12.7h)  &  12.86& 13.37     \\ 
\bottomrule
\end{tabular}
\label{SOTA}
\vspace{-0.4cm}
\end{table}

\vspace{-0.4cm}
\subsection{Comparison with the State-of-the-Art}
\vspace{-0.2cm}
We compare the proposed self-training solutions with the existing supervised models on Kaldi-hybrid networks~\cite{dabike2019automatic,demirel2020automatic,ahlback2021mstre} and E2E frameworks~\cite{gao2022automatic,ou2022towards} for lyrics transcription of solo-singing in Table.~\ref{SOTA}. We observe that our Self-Transriber models trained on 12.7 hours' labeled data outperform most supervised models~\cite{dabike2019automatic,demirel2020automatic,ahlback2021mstre,gao2022automatic} trained on 149.1 hours' labeled data, and achieves competitive performance compared with~\cite{ou2022towards} trained on 149.1 hours' labeled data. A demo page with examples of this work is available in the link~\footnote{\url{https://xiaoxue1117.github.io/icassp2023/}}.

\vspace{-0.4cm}
\section{Conclusions}
\vspace{-0.4cm}
This work enables lyrics transcription with as little as 12.7 hours labeled data with 13.37 \% WER that outperforms the best system (13.46 \% WER) four month ago trained on 149.1 hours of labeled data. The effectiveness of noisy student augmentation for unsupervised data is also verified via experiments. To this regard, our Self-Transcriber has proven that the lyrics transcription can be achieved with very small amount of labeled data at a good accuracy. Self-training of lyrics transcription is firstly presented with different amount of unlabeled data, which provides insights for lyrics transcription towards data-efficient semi-supervised learning.
This work also makes lyrics transcription more flexible and easier to be adapted to many other languages that are difficult and expensive to obtain hundreds of hours of labeled singing data. We would like to explore unsupervised pre-training and self-training together for lyrics transcription in the future.

\footnotesize
\textbf{Acknowledgement}
This work is supported by 1) the Agency for Science, Technology and Research (A*STAR) under its AME Programmatic Funding Scheme (Project No. A18A2b0046) and 2) IAF, A*STAR, SOITEC, NXP and National University of Singapore under FD-fAbrICS: Joint Lab for FD-SOI Always-on Intelligent \& Connected Systems (Award I2001E0053). We also thank Dr. Junichi Yamagishi's contribution for this work.

\vfill\pagebreak

\footnotesize
\bibliographystyle{IEEEtran}
\bibliography{strings}

\end{document}